\documentclass[twocolumn,prl,showpacs]{revtex4}
\pdfoutput=1   
\usepackage{graphicx}
\usepackage{rotating}
\usepackage{amsmath}
\usepackage{amsfonts}
\usepackage{amssymb}
\usepackage{enumerate}
\usepackage{longtable}
\usepackage{dcolumn} 
\usepackage{bm}
\usepackage{float}

\begin{document}

\newcommand{\ii}{\'{\i}}

\title{Mott-Kondo Insulator Behavior in the Iron Oxychalcogenides}

\author{B. Freelon,$^{1,2*}$  Yu Hao Liu,$^{1}$ Jeng-Lung Chen,$^{1,3}$ L. Craco,$^{4*}$ M. S. Laad,$^{5}$ S. Leoni,$^{6}$ 
Jiaqi Chen,$^{7}$ Li Tao,$^{7}$ Hangdong Wang,$^{7}$ R. Flauca,$^{8}$ Z. Yamani,$^{8}$ Minghu Fang,$^{7}$ Chinglin Chang,$^{3}$ J.-H. Guo$^{1}$ and Z. Hussain$^{1}$}

\email{freelon@mit.edu, lcraco@fisica.ufmt.br}

\affiliation{$^1$ Advanced Light Source Division, Lawrence Berkeley National 
Laboratory, One Cyclotron Road, Berkeley, CA 94720 \\
$^2$Advanced Photon Source, Argonne National Laboratory, Argonne, Illinois 60439\\
$^3$Department of Physics, Tamkang University, Tamsui, Taiwan 250 \\
$^4$Instituto de F\ii sica, Universidade Federal de 
Mato Grosso, 78060-900, Cuiab\'a, MT, Brazil   \\
$^5$The Institute of Mathematical Sciences, C.I.T. Campus, 
Chennai 600 113, India \\
$^6$School of Chemistry, Cardiff University, Cardiff, CF10 3AT, UK\\
$^7$Department of Physics, Zhejiang University, Hangzhou 310027, P. R. China \\
$^8$Canadian Neutron Beam Centre, National Research Council, Chalk River Laboratories, Chalk River, Ontario K0J 1J0, Canada
}

\date{\today}

\begin{abstract}
We perform a combined experimental-theoretical study of the Fe-oxychalcogenides (FeO$\emph{Ch}$) series 
La$_{2}$O$_{2}$Fe$_{2}$O\emph{M}$_{2}$  (\emph{M}=S, Se), which is the latest among the Fe-based materials with the potential \
to show unconventional high-T$_{c}$ superconductivity (HTSC). A combination of incoherent Hubbard features
in X-ray absorption (XAS) and resonant inelastic X-ray 
scattering (RIXS) spectra, as well as resitivity data, reveal that the parent 
FeO$\emph{Ch}$ are correlation-driven insulators. To uncover microscopics underlying these findings, we perform local density 
approximation-plus-dynamical mean field theory (LDA+DMFT) calculations that 
unravel a Mott-Kondo insulating state. Based upon good agreement between theory and a range of data, we propose that FeO$\emph{Ch}$ may constitute a new, ideal testing ground to explore HTSC arising from a strange metal proximate to a novel selective-Mott quantum criticality. 

\end{abstract}


\maketitle

Since the discovery~\cite{doi:10.1021/ja800073m} of high-temperature 
superconductivity (HTSC) in iron pnictides (FePn), a fundamental question 
has been whether the iron pnictides are weakly correlated metals or 
bad metals in close proximity to Mott 
localization~\cite{PhysRevLett.101.076401,Physics.1.21}. These competing 
ideas must imply very different mechanisms of Fe-based HTSC. The Mottness view 
proposes that strong, orbitally-selective electron correlation is relevant 
to HTSC, while the competing itinerant view holds that antiferromagnetic (AFM) order 
and superconductivity (SC) originate from a Kohn-Luttinger-like Fermi surface 
(FS) nesting mechanism with weak electronic 
correlations ~\cite{PhysRevB.78.134512}. Early insights suggested the 
importance of varying the 
strength~\cite{GBaskaranJPSJ2008,PhysRevLett.101.076401} of electron 
correlations using chemical substitutions to tune systems across band-width 
driven Mott localization.  Since then, more evidence has accumulated in 
support of this proposal; FeSe is a bad-metal without AF 
order~\cite{PhysRevLett.102.177005}, while parent Fe-oxychalcogenides are 
electrical insulators ~\cite{PhysRevLett.104.216405,0953-8984-26-14-145602}.   

Most FePn superconductors exhibit bad-metal normal states 
with hints of novel quantum criticality. 
The discovery of HTSC in  alkaline iron selenides 
\emph{A}$_{1-x}$Fe$_{2-y}$Se$_{2}$, where \emph{A} = K, Tl, Cs and Rb (referred 
to as ''245s'') intensified this basic debate, since SC with T$_{c}$ = 30~K emerges from doping an AFM insulating state occuring due to Fe-vacancy induced band narrowing~\cite{0295-5075-94-2-27009}. 
Studying more such materials to further develop a complete description of the generic electronic phase 
diagram of iron superconductors is clearly warranted. With this in mind, 
we investigated iron oxychalcogenides such as 
\emph{X}$_{2}$OFe$_{2}$O$_{2}$\emph{M}$_{2}$ [\emph{X}=(La, Na, Ba), 
(\emph{M}=(S, Se)] ~\cite{ANIEBACK:ANIE199216451} because of $(i)$ their 
similarities to LaFeOAs (1111) materials and $(ii)$ the opportunity to 
study materials with slightly larger electron correlation than the 
pnicitides~\cite{PhysRevLett.104.216405}.  Moreover, 
La$_{2}$O$_{2}$Fe$_{2}$O\emph{M}$_{2}$  (\emph{M}=S, Se) offer the 
attractive option to tune the strength of electronic correlations in 
one system through disorder-free chalcogen replacement of S by Se.  The FeO\emph{Ch} 
consist of stacked double-layered units La$_{2}$O$_{2}$ and 
Fe$_{2}$O(S, Se)$_{2}$. Fe$_{2}$O(S, Se)$_{2}$ contains an Fe layer that 
is similar to the FeAs layer in the pnictides; in the latter, Fe$^{2+}$ 
($d^{6}$ configuration) is co-ordinated by O atoms but there is an 
extra Oxygen atom O(2). The rare-earth layer also contains an additional 
O(1) atom. Chalcogen substitution results in a slightly expanded 
Fe-square-lattice unit cell of La$_{2}$O$_{2}$Fe$_{2}$O\emph{M}$_{2}$ 
compared to that of LaOFeAs~\cite{ANIEBACK:ANIE199216451}. The consequent 
decrease in the electronic bandwidth \emph{W} must increase the Hubbard 
\emph{U}, offering a 
disorder-free parameter to subtly tune \emph{U/W} across a critical 
value for a Mott insulator phase.  

In this Letter, we substantiate this reasoning by presenting first
measurements of the unoccupied and occupied Fe and Oxygen density-of-states 
(DOS) obtained by soft X-ray absorption (XAS) and resonant inelastic X-ray 
scattering (RIXS). Strong Fe moment localization that can be tuned 
with chalcogen replacement, and, more crucially, the observation of incoherent Fe spectral weight 
due to Hubbard band formation marks the FeO$Ch$ as correlation-driven insulators. 
The incoherent spectral weight undergoes a resonant enhancement in the RIXS 
data and LHB spectral weight analysis, indicating 
\begin{figure}[t]
	\vspace{-21.0em}
      \includegraphics[width=3.7in]{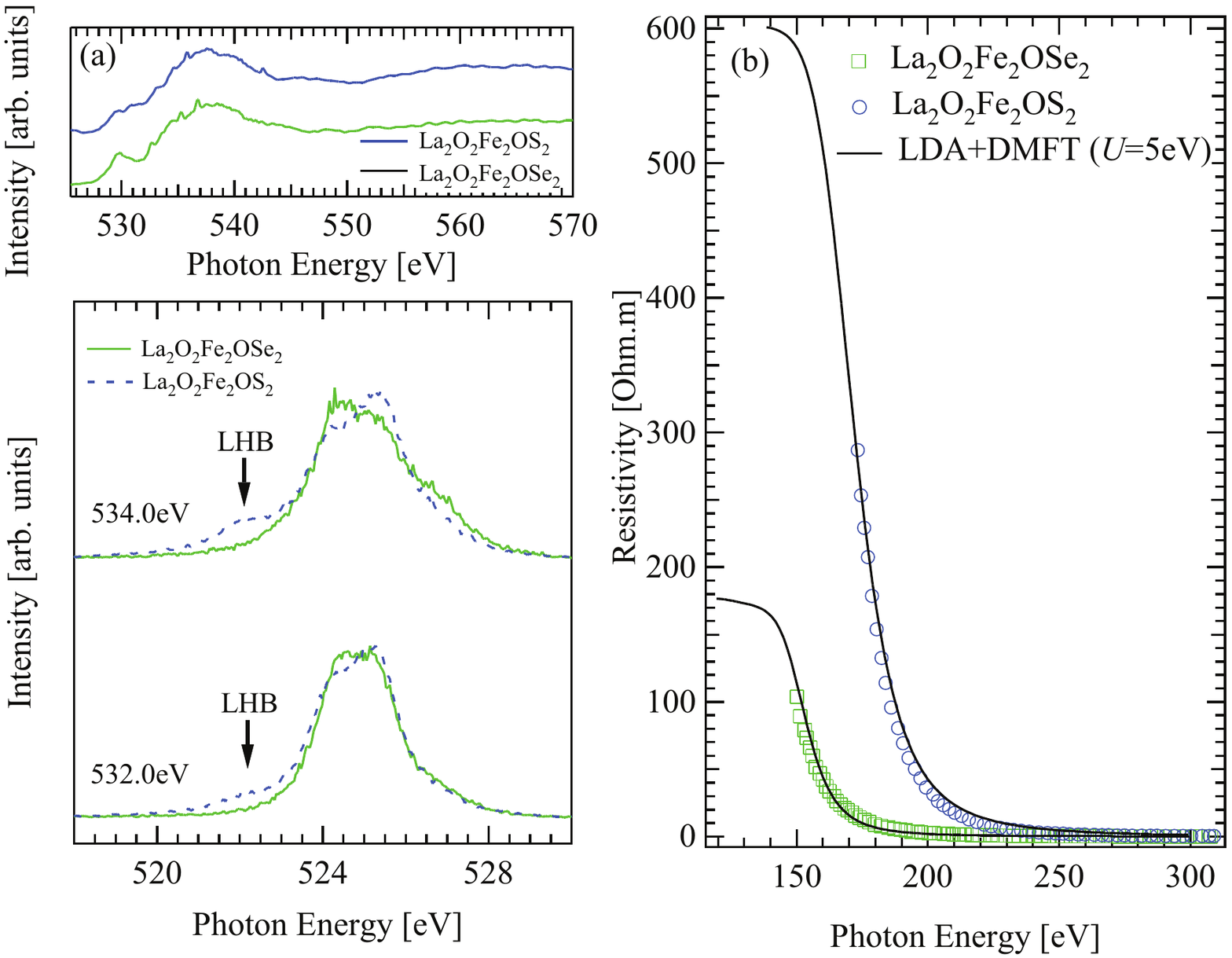}
      \caption{(Color online) (a) O 1$\emph{s}$ XAS  and (b) O $\emph{K}\alpha$ 
RIXS data for La$_{2}$O$_{2}$Fe$_{2}$O\emph{M}$_{2}$, (\emph{M} = S, Se). 
RIXS intensity data was collected using photons with incident energies of 
532 and 534 eV. In (c) the electrical resistivity versus temperature \emph{T} data
and the simulated resistivity calculated using LDA + DMFT is shown.}
\label{fig1}
    \end{figure} the Mott insulating character 
of both La$_{2}$O$_{2}$Fe$_{2}$OS$_{2}$ and La$_{2}$O$_{2}$Fe$_{2}$OSe$_{2}$. We 
discuss the implications of these findings for possible HTSC arising from an incoherent 
metal proximate to novel selective-Mott quantum criticality.

We investigated well characterized La$_{2}$O$_{2}$Fe$_{2}$O\emph{M}$_{2}$  
\emph{M}=(S, Se) polycrystalline materials, with nominal compositions, 
that were fabricated using solid state reactions  methods described 
earlier~\cite{PhysRevLett.104.216405}. High purity La$_{2}$O$_{3}$, Fe 
and (S, Se) powders were used as starting materials and laboratory 
X-ray powder diffraction data showed that these materials had minimal 
impurity phases; neutron powder diffraction data confirmed this finding.  

XAS and RIXS spectra for the FeO$Ch$ were collected in a polarization 
averaged condition. In order to reduce the possibility of 
oxidation, iron oxychalcogenide samples were sheared under a pressure 
of 10$^{-6 }$ Torr in a pre-chamber immediately before being placed in 
the ultra-high vacuum experimental chamber. All presentend 
measurements were performed in the paramagnetic 
state above 150 K. The Advanced Light Source beamlines 
7.0.1 and 8.0.1 delivered X-ray beams of 100 micron spot-sizes and energy 
resolutions of 0.2~eV~(0.2~eV) and 0.5~eV~(0.6~eV) for oxygen(iron) X-ray 
absorption and emission, respectively. The oxygen \emph{K}-edge XAS 
intensity is proportional to the unoccupied \emph{p}-weighted DOS 
resulting from the $\Delta{\emph{l}}$ = $\pm{1}$ dipole selection rule 
for X-ray photon absorption. If the Fe-O related bond is partially covalent, 
O $2p$ states hybridize with Fe $3d$ orbitals and hence the pre-threshold 
for absorption, i.e., the low energy fine structure, reveals itself in 
ligand-to-metal charge transfer excitations~\cite{PhysRevB.40.5715}. XAS 
is sensitive to metal atom coordination geometry and $3d$ occupation 
(oxidation state).  \begin{figure}[t]
	\vspace{-20.0em}
\includegraphics[width=3.4in]{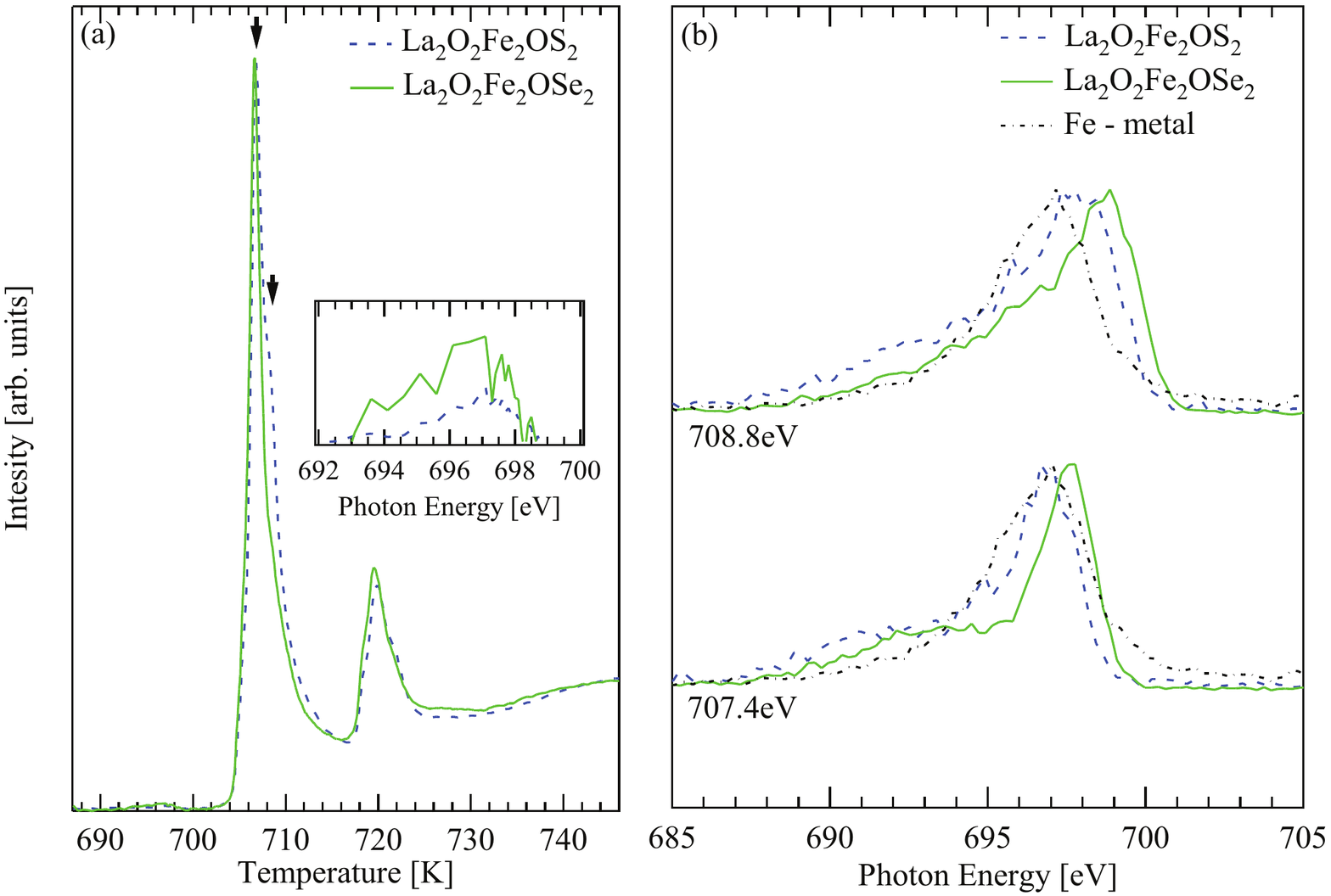}
\caption{(Color online) 
The Fe \emph{L}$_{2,3}$-edge XAS profiles of 
La$_{2}$O$_{2}$Fe$_{2}$O\emph{M}$_{2}$, (\emph{M} = S, Se) is shown in panel 
(a). The arrows indicate the X-ray absorption features at the photon 
energies, 707.4  and  708.8 eV, used to collect the Fe RIXS data in (b).  
The data are plotted with a vertical offset for clarity.}
\label{fig2}
\end{figure} Fig.~\ref{fig1}, top panel~(a), shows the O 
\emph{K} XAS profiles of La$_{2}$O$_{2}$Fe$_{2}$O\emph{M}$_{2}$ 
(\emph{M} = S,~Se). Incident X-ray energies of 532 and 534~eV were 
used to collect the O $\emph{K}\alpha$ RIXS data presented in 
Fig.~\ref{fig1}~(a). The presence of lower Hubbard band (LHB) structure 
in La$_{2}$O$_{2}$Fe$_{2}$OS$_{2}$ RIXS spectra is consistent with resistivity 
data (Fig.~\ref{fig1}~(b)) which shows \emph{M} = S to be a better insulator 
than \emph{M} = Se.  We note the absence of O charge transfer band, i.e., 
a high energy shoulder structure for the O $2p$ mainband peak, in the 
Oxygen RIXS data of La$_{2}$O$_{2}$Fe$_{2}$OSe$_{2}$ material: this suggests that 
the extent of Fe-O hybridization DOS is different for \emph{M} = S and 
\emph{M}=Se compounds.

In Fig.~\ref{fig2}~(a), we show the Fe \emph{L$_{2,3}$}-edge XAS spectra 
for both La$_{2}$O$_{2}$Fe$_{2}$OS$_{2}$ and La$_{2}$O$_{2}$Fe$_{2}$OSe$_{2}$. 
A magnification (inset) of the region near the Fermi level, presented in 
Fig.~\ref{fig2}~(a), reveal a clear difference in the iron conduction 
band weight of the two materials. La$_{2}$O$_{2}$Fe$_{2}$OSe$_{2}$ has 
enhanced spectral weight at pre-edge just above $E_{F}$, and this must ultimately  
be related to details of the {\it correlated} electronic structure 
changes upon replacement of S by Se.  Fe XAS data is consistent with both RIXS and 
bulk measurements that show \emph{M} = Se to be less strongly correlated 
than \emph{M} = S. 
Resonant inelastic X-ray 
scattering (RIXS) intensity data, see Fig.~\ref{fig2}~(b), were collected 
using incident X-ray energies tuned to near- and off-resonance Fe X-ray 
absorption features at 707.4~eV and 708.8~eV, respectively.  
RIXS~\cite{RevModPhys.73.203} intensity results from a second-order (i.e., 
two-step) photon-in, photon-out process that can be can be described by 
the Kramers-Heisenberg (KH) relationship~\cite{RevModPhys.73.203}. The 
KH expression requires the enhancement of the scattering intensity when 
incident photon energy matches X-ray absorption edge energies. The RIXS 
data for \emph{M} = S shows correlation induced broad features in the 
high energy spectral region that form the LHB. In the RIXS data, for both
\begin{figure}[t]
	\vspace{-22.0em}
\includegraphics[width=3.5in]{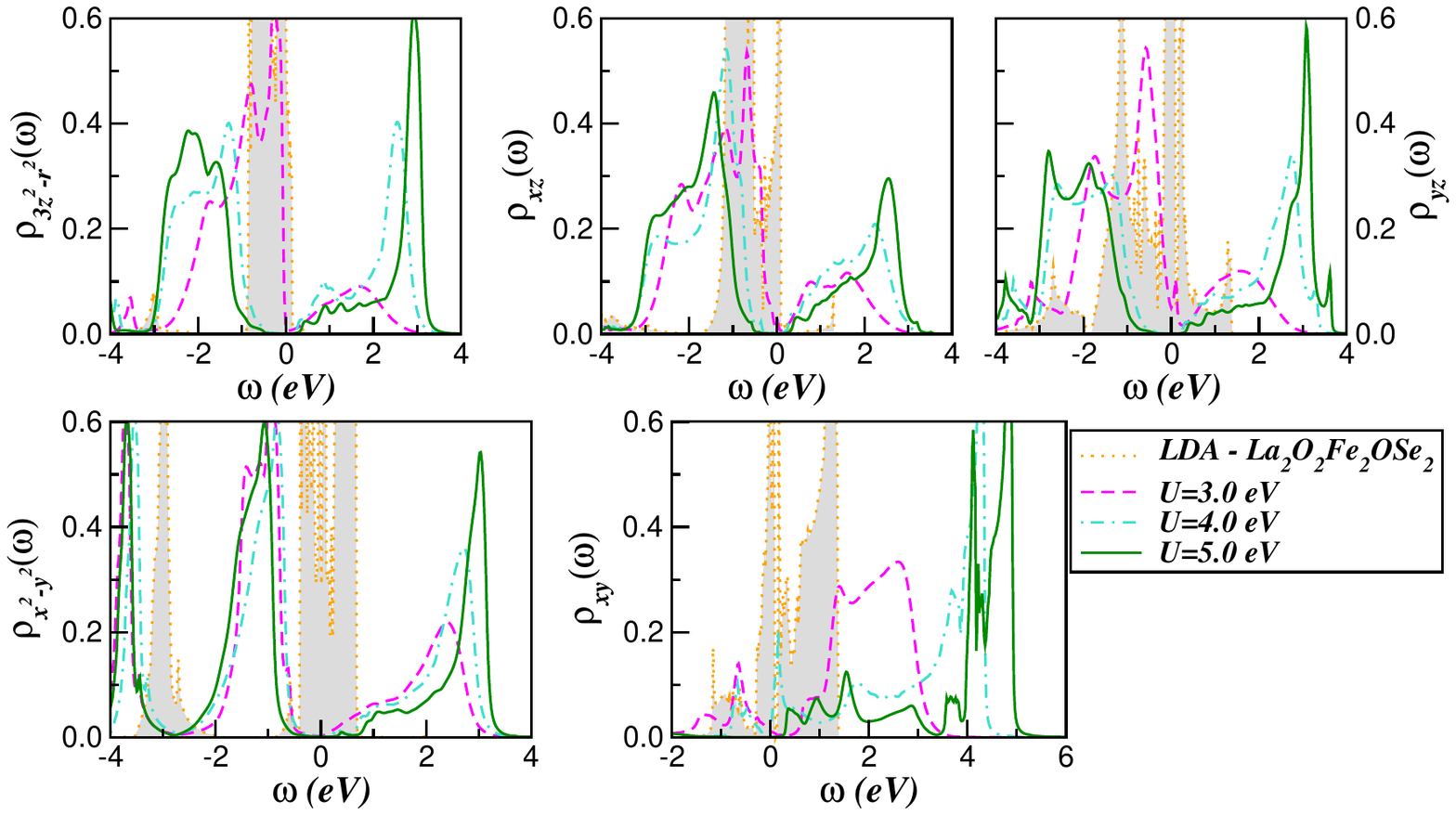}
\caption{(Color online)
Orbital-resolved LDA  density-of-states (DOS) for the Fe $d$ orbitals of
La$_{2}$O$_{2}$Fe$_{2}$OSe$_{2}$ as well as LDA+DMFT results for different values 
of $U$ (with $U'=U-2J_H$) and fixed $J_{H}=0.7$~eV. Notice the narrow 
bands in the LDA DOS. Compared to the LDA results, large spectral weight 
transfer along with electronic localization is visible in the LDA+DMFT 
spectral functions for $U \geq 4$~eV.}
\label{figTheo1}
\end{figure} \emph{M} = S,~Se, there is a high intensity Fe $3d$ metal valence band 
peak. Near 692~eV, a broad shoulder in the high energy spectral region is 
resonantly enhanced. The spectra clearly reveal incoherent electronic 
excitations involving $3d$ Fe electrons. These are correlation satellites 
formed by multi-particle excitations that involve localized states, and clearly
reveal the stronger electronic correlations that generate a correlated 
insulator in the FeO$\emph{Ch}$.

Taken together, these results mark
the FeO$\emph{Ch}$ as correlated, 
bandwidth-control driven insulators~\cite{RevModPhys.70.1039}.  In multi-orbital FeO$\emph{Ch}$ with strong inter-orbital charge transfer, correlations and crystal field effects, proper characterization of the Mott insulating 
state(s) is an important, non-trivial issue. To shed light on microscopics underlying the above findings, we performed LDA+DMFT calculations for the FeO$\emph{Ch}$ closely following 
earlier work~\cite{0953-8984-26-14-145602,PhysRevB.84.224520}.  LDA 
calculations were performed within the linear muffin-tin orbitals 
(LMTO) scheme.  The one-electron Hamiltonian reads 
$H_{0}=\sum_{k,a,\sigma}\epsilon_{a}(k)c_{k,a,\sigma}^{\dag}c_{k,a,\sigma}$ with 
$a=xy,xz,yz,x^{2}-y^{2},3z^{2}-r^{2}$ label (diagonalized in orbital basis) 
five Fe $3d$ bands.  We retain only the $d$ states, since the non-$d$ 
orbital DOS have negligible or no weight at $E_{F}$.  An important aspect 
of our study is readily visible in Fig.~\ref{figTheo1} which shows a sizable 
reduction of the LDA bandwidth relative to that of FeSe (which is itself 
an incoherent bad-metal~\cite{0295-5075-91-2-27001}),
induced by hybridization with $O$ states and distorting effects of the 
La-layers.
 Another interesting aspect, in contrast to Fe-arsenides, is that 
the $xz-yz$ orbital degeneracy in the tetragonal structure is explicitly 
removed at the outset. Thus, in FeO$\emph{Ch}$, an electronic nematic (EN) 
instability linked to ferro-orbital order and its potentially related criticality 
is not relevant.  Finally, clear lower-dimensional band structural features 
are visible in $\rho_{LDA}^{xz}(\omega)$ (quasi-1D) and $\rho_{LDA}^{xy}(\omega)$ 
(2D van-Hove singular) near the Fermi energy, $E_{F}(=0)$.  Enhancement of 
the {\it effective} $U/W$ ratio in such an intrinsically anisotropic setting 
should now naturally favor the Mott insulator state in FeO$\emph{Ch}$. To 
substantiate this, DMFT calculations were performed using the multi-orbital 
iterated perturbation theory (MO-IPT) as an impurity solver. Though not 
exact, MO-IPT has a proven record of providing reliable results for a 
number of correlated systems~\cite{RevModPhys.68.13}, and is a very 
efficient and fast solver for arbitrary $T$ and band-filling. The full 
Hamiltonian is $H=H_{0}+H_{1}$ with $H_{1}= U\sum_{i,a}n_{ia\downarrow}n_{ia\uparrow} + \sum_{i,a\neq b}[U'n_{ia}n_{ib}-J_{H}{\bf S}_{ia} \cdot {\bf S}_{ib}]$, with 
$U' = (U-2J_{H})$ being the inter-orbital Coulomb term and $J_{H}$ the 
local Hund's coupling. Though $U\simeq 4.0-5.0$~eV and $J_{H}=0.7$~eV 
are expected, we vary $3.0\leq U \leq 5.0$ to get more insight.  

\begin{figure}[t]
	\vspace{-22.0em}

\includegraphics[width=3.5in]{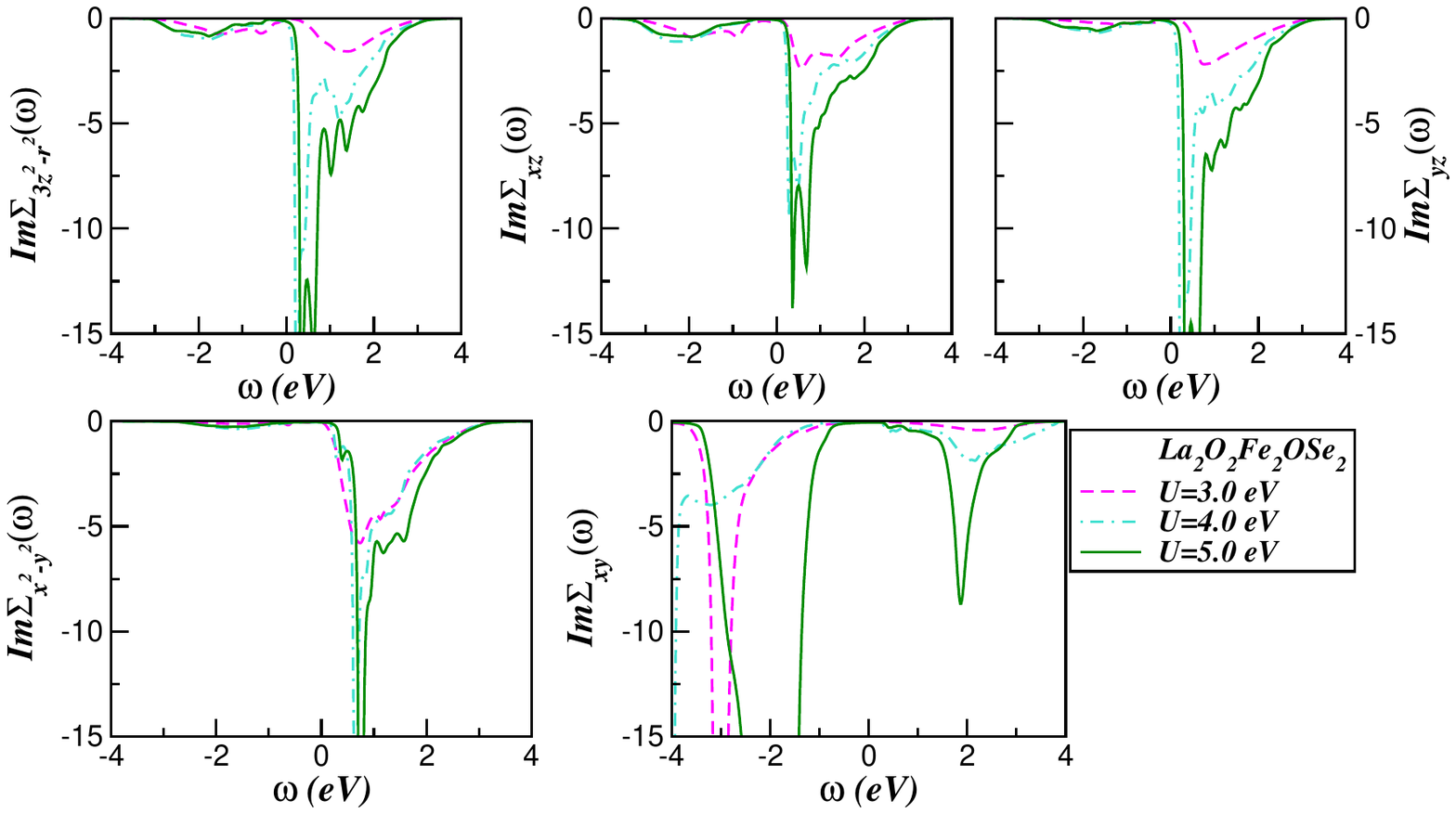}

\caption{(Color online)
Orbital-resolved imaginary parts of the DMFT self-energies for the 
Fe $d$ orbitals of La$_{2}$O$_{2}$Fe$_{2}$OSe$_{2}$ for different 
values of $U$ and $J_{H}=0.7$~eV.}
\label{figTheo2}
\end{figure}

In Fig.~\ref{figTheo1}, we show the total many-body local spectral function 
(DOS) for the $\emph{M}$ = Se case. A clear transition from a bad-metal 
state at $U=3.0$~eV to a correlated insulator for $U=4.0,~5.0$~eV occurs.  
Looking closely at the self-energies in Fig.~\ref{figTheo2}, however, a 
remarkable aspect

stands out: Im$\Sigma_{a}(\omega=E_{F})$ vanishes in 
the region of the Mott gap, instead of having a pole, as would occur 
in a true Mott insulator.  This aspect is more reminiscent of a correlated 
{\it Kondo} insulator, where the gap arises due to combined effects of 
electronic correlations and sizable interband hybridization. While 
this novel finding is somewhat unexpected, proximity to a true Mott state 
is seen by observing that the pole structures in Im$\Sigma_{a}(\omega)$ 
with $a=xz,yz,3z^{2}-r^{2}$ lie only slightly above $E_{F}$.  Thus, because 
the system {\it is} correlated, sizable spectral redistribution in response 
to additional small perturbations (e.g, longer-range Coulomb interactions in an insulator) 
can drive it into a true Mott insulating phase. More importantly, electron 
doping will also generically cause large-scale orbital-selective (OS) spectral 
weight transfer,
\begin{figure}[t]
	\vspace{-17.0em}

\includegraphics[width=3.3in]{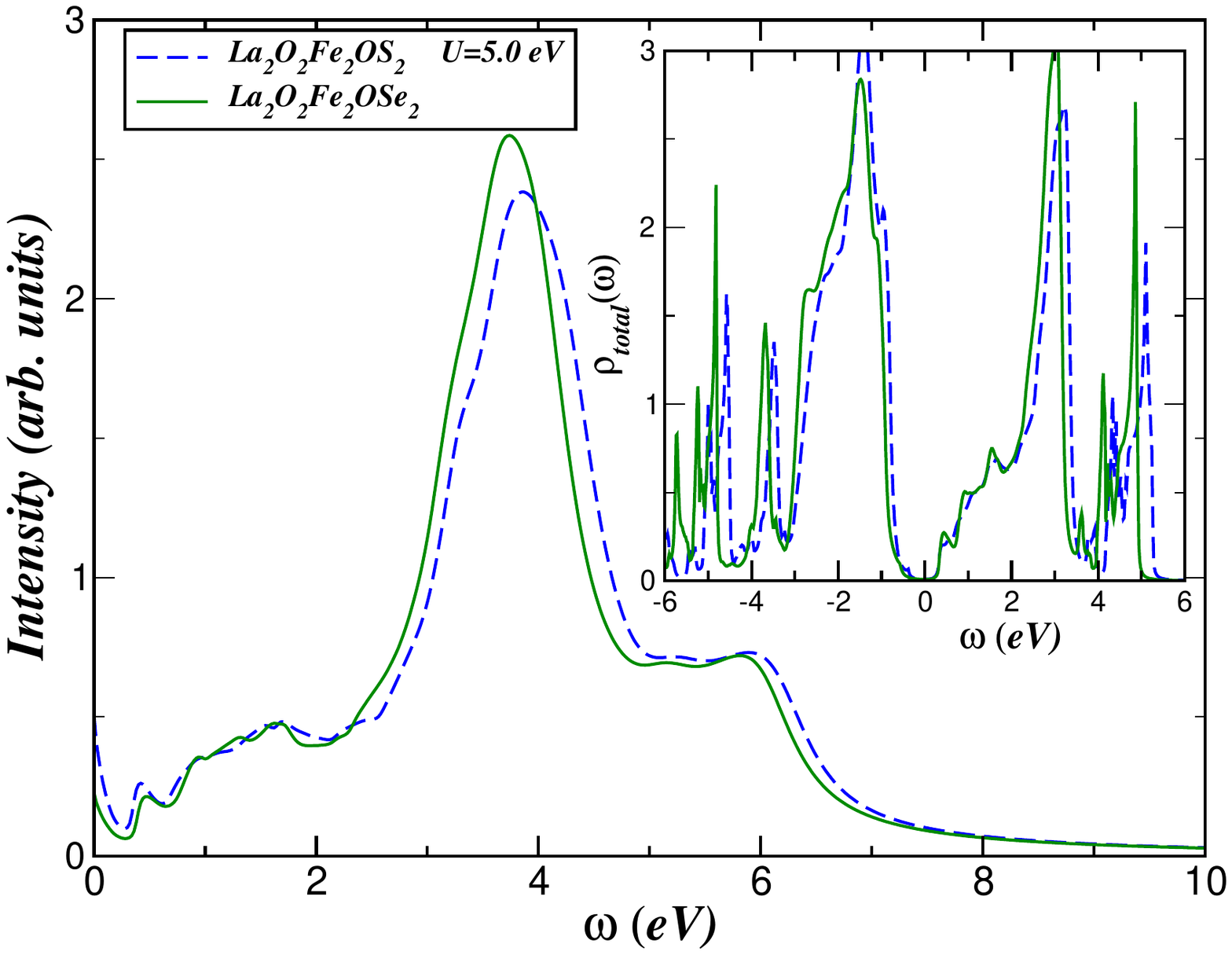}

\caption{(Color online)
Total LDA+DMFT XAS spectrum of La$_{2}$OFe$_{2}$O$_{2}M_{2}$ $(M=$S,~Se), 
showing the changes in the upper Hubbard band upon chalcogen substitution. 
Inset shows the corresponding total LDA+DMFT spectral function for 
$U=$5.0~eV and $J_H=$0.7~eV.}
\label{figTheo3}
\end{figure} leading to orbital-selective Mott transition (OSMT), 
wherein a subset of orbital states remain (Mott) gapped in an incoherent 
metallic state~\cite{0953-8984-26-14-145602}. Such states can exhibit 
novel quantum criticality associated with an end-point of the OSMT.  
Whether HTSC near such novel criticality occurs in 
suitably doped FeO$\emph{Ch}$ is an enticing and open issue.

Remarkably, a comparison to Fe-edge XAS and $dc$ resistivity 
data reveals good agreement, providing support for this novel finding.  
In Fig.~\ref{figTheo3}, we show our LDA+DMFT XAS spectrum. We have incorporated the relevant effects of core-hole scattering for XAS lineshapes by consistently adapting the procedure of Pardini {\it et al.}~\cite{0953-8984-23-21-215601} within our DMFT scheme.  Several important aspects stand out: $(i)$ there is good semi-quantitative accord with 
with the Fe-edge XAS data up to $3.0$~eV above $E_{F}$. More importantly, 
details of the {\it differences} in spectral weight between $M=$S and $M=$Se 
cases is also in very good accord with data, $(ii)$ enhanced spectral weight 
at pre-edge just above $E_{F}$, just as seen, while smearing of lower energy 
structures ($0.0<\omega<2.0$~eV) and higher energy of the $3.0$~eV peak for 
$M=$S relative to $M=$Se is in good accord, both with XAS data, as well as 
with resistivity data which show $M=$S to be more correlated than $M=$Se.  
Indeed, LDA+DMFT resistivities (using 
the Kubo formalism and neglecting vertex corrections to the conductivity, 
which are negligible~\cite{0953-8984-21-6-064209}) for $M=$S, Se show very good accord with 
experiment (where $\rho_{dc}^{(M)}(T)$ was measured above $150$~K. 
This is a strong internal self-consistency check, and puts our novel proposal on stronger ground.

  Armed with these positive features, we now discuss their implications for possible HTSC
in suitably doped or pressurized systems.  Both are expected to tune the Fe-O hybridization: since we
find a correlated Mott-Kondo insulator, this must non-trivially reconstruct the Fe-$d$ states. 
In La$_{2}$O$_{2}$Fe$_{2}$OSe$_{2}$, suppression of oxygen Hubbard-band spectral 
weight presages the onset of bad-metallicity even as the localization of 
the Fe $3d$ states is preserved. This may provide clues to metallize FeO$Ch$ 
by appropriately manipulating the Oxygen and chalcogen states which hybridize 
with Fe-$d$ states. This is an exciting perspective because, if HTSC results,
(as in 245s), it would reinforce the fundamental  challenge to the itinerancy 
perspective in pnictide HTSC research. Specifically, since strong correlations 
have now obliterated the LDA FS as above, large-scale orbital-selective redistribution 
of spectral weight upon carrier doping is generically expected to lead to OSMT 
and FS involving only a subset of original $d$-orbitals. 
The LDA FS and its nesting features can now no longer serve as a guide for 
rationalizing instabilities to novel order(s). Instead, strong electronic 
scattering between the partially (Mott) localized and itinerant subsets of 
the many-body spectral functions generically extinguishes the Landau 
quasiparticle (QP) pole at $E_{F}$.  In absence of a QP description for 
the normal state, the very tenability of the FS nesting scenario is called 
into doubt.

Thus, our work provides an impetus to consider closer similarities between 
Fe-based systems and the cuprates.

In similarity with 245 materials, we suggest that FeO$\emph{Ch}$ are ideal 
candidates for testing this proposal. The development of incoherent 
scattering observed in our XAS and RIXS spectra arises from Mott localized 
states. If these can be appropriately doped, it is likely that 
strong orbitally-selective scattering will nearly extinguish the Landau-Fermi 
liquid quasiparticles, leading to the emergence of an incoherent pseudogapped 
metal reminiscent of underdoped cuprates~\cite{0953-8984-26-14-145602}. 
Whether novel quantum criticality associated with a selective-Mott transition 
in conjunction with the development of HTSC in FeO$\emph{Ch}$  will be seen 
in the future is of great interest. Our findings provide a compelling motivation 
to produce and study doped or pressurized iron oxychalcogenides.  

After the recent journal submission of this paper, we learned of work by G. Giovannetti \emph{et. al.} on the iron oxychalcogenides which is complementary to ours and gives comparable results.

\begin{acknowledgments}
The Advanced Light Source is supported by the Director, Office of Science, 
Office of Basic Energy Sciences, of the U.S. Department of Energy (DOE), 
under Contract No. DE-AC02-05CH11231. ZJU work is supported by the National 
Basic Research Program of China (973 Program) under Grants No. 2011CBA00103 
and 2012CB821404, the National Science Foundation of China (No. 11374261, 
11204059) and Zhejiang Provincial Natural Science Foundation of China 
(No. LQ12A04007). M.S.L thanks MPIPKS, Dresden for hospitality.  
L.C.'s work was supported by CAPES - Proc. No. 002/2012. 
Acknowledgment (L.C.) is also made to FAPEMAT/CNPq (Project: 685524/2010) 
and the Physical Chemistry department at Technical University Dresden for 
hospitality. 
\end{acknowledgments}

\bibliography{freelonRef1st_refcheck}


\end{document}